\documentclass[conference]{IEEEtran}
\IEEEoverridecommandlockouts
\usepackage{cite}
\usepackage{amsmath,amssymb,amsfonts}
\usepackage{algorithmic}
\usepackage{graphicx}
\usepackage{textcomp}
\usepackage{xcolor}
\usepackage{wrapfig}
\usepackage{placeins}

\def\BibTeX{{\rm B\kern-.05em{\sc i\kern-.025em b}\kern-.08em
    T\kern-.1667em\lower.7ex\hbox{E}\kern-.125emX}}
\begin{document}

\author{
\IEEEauthorblockN{Dexter Burns}
\IEEEauthorblockA{
    University of Florida\\
    Department of Electrical and Computer Engineering\\
    Gainesville, FL, USA\\
    burns.dexter@ufl.edu
}
\and
\IEEEauthorblockN{Dr Sanjeev Koppal}
\IEEEauthorblockA{
    University of Florida\\
    Department of Electrical and Computer Engineering\\
    Gainesville, FL, USA\\
    sjkoppal@ece.ufl.edu
}
}

\title{Active Optics for Hyperspectral Imaging of Reflective Agricultural Leaf Sensors}
\maketitle

\begin{abstract}
Monitoring plant health increasingly relies on leaf-mounted sensors that provide real-time physiological data, yet efficiently locating and sampling these sensors in complex agricultural environments remains a major challenge. We present an integrated, adaptive, and scalable system that autonomously detects and interrogates plant sensors using a coordinated suite of low-cost optical components including a LiDAR, liquid lens, monochrome camera, filter wheel, and Fast Steering Mirror (FSM). The system first uses LiDAR to identify the distinct reflective signatures of sensors within the field, then dynamically redirects the camera s field of view via the FSM to target each sensor for hyperspectral imaging. The liquid lens continuously adjusts focus to maintain image sharpness across varying depths, enabling precise spectral measurements. We validated the system in controlled indoor experiments, demonstrating accurate detection and tracking of reflective plant sensors and successful acquisition of their spectral data. To our knowledge, no other system currently integrates these sensing and optical modalities for agricultural monitoring. This work establishes a foundation for adaptive, low-cost, and automated plant sensor interrogation, representing a significant step toward scalable, real-time plant health monitoring in precision agriculture.
\end{abstract}


\begin{figure}
    \centering
    \includegraphics[width=0.75\linewidth]{Images/System_Labeled.jpg}
    \caption{\centering\scriptsize Full System: 1 – Monochrome Camera; 2 – Focal Lens; 3 – Liquid Lens; 4 – Filter Wheel; 5 – Fast Steering Mirror (FSM); 6 – Velodyne HDL-32 LiDAR}
    \label{fig:full_system}
\end{figure}

\section{Introduction and Background}
There are many initiatives to integrate technology into agriculture such as IoT4Ag, Ceres2030, and Agritech4.0 where many researchers, professors, professionals, and students have come together to create and invest in technologies that further engineering research in agriculture with the hopes that the projects and inventions of the initiative will make agriculture more efficient for agricultural farmers \cite{IoT4Ag2021, Ceres20302021, AgriTech2022}. One of the main visions of the initiative is to transform agriculture through the deployment of networks of miniature, low-cost sensors embedded directly within the agricultural environment. These tiny leaf/stem-mounted sensors \cite{Kuruppuarachchi2025WearableSensors} measure micro-scale variables such as water content, nutrient availability, or stress markers, providing unprecedented insight into plant health and field conditions. However, we are limited by the difficulty of locating and sampling these sensors in agricultural contexts. In particular, their small size, passive design and spatial distribution make traditional collection methods inefficient or cost prohibitive. We address these limitations in our work and successfully surmount those problems. 

We recognize that the implications of our work advances the feasibility of fully automated, detailed plant monitoring. Such capability has significant global implications. Automated sensing systems such as ours can drastically improve the efficiency of agricultural management by reducing input waste, optimizing irrigation and fertilization and enabling early access of plant stress or disease \cite{Cheng2023, Oliveira2021, Shamshiri2018}. These improvements directly translate to higher yield, more resilient crops and more sustainable land use. More importantly, as the global population increases and as climate change places new stresses on food systems, scalable monitoring technologies are essential to meet rising food demand, and our methodology contributes to the broader goal of ensuring food security with precision agriculture sensing methods that help close the gap between current production and future needs \cite{FoodInsecurity2018, 10927238}. 

The sensors we are detecting work by modulating their hyperspectral signature in response to the physiological changes of the plant that it is currently mounted on \cite{Kramadhati2025-PlantSensors, Mallavarapu2023-Metasurfaces, Guo2012-Biosensors}. Depending on their design and fabrication, these sensors may respond to variables such as water content or stress markers. In all cases, their hyperspectral response encodes plant state information. \cite{Thalheimer2023} These sensors have to be imaged closely and in the hyperspectral range so that we can properly evaluate the state of the sensor, and thus the state of the plant. Current hyperspectral solutions are also expensive and not well suited for scalable field deployment \cite{Bhargava2024HSIReview}. 

Our solution surmounts these aforementioned challenges with our integrated sensing pipeline that we have developed. The system first utilizes a LiDAR for plant sensor acquisition. It locates the plant sensor's 3D location in space and isolates the sensor from the rest of the environment by filtering for the unique reflectivity range of the plant sensors \cite{WeissBiber2011}. Once the sensor has been found, we get an estimate on how far that sensor is from the system, which then prompts our liquid lens to continually adjust itself so that the plant sensor remains in sharp focus for the camera. Simultaneously, the FSM adjusts to redirect the FOV of the camera to align the camera FOV with the sensor's position \cite{FastSteeringMirror}. From there, the monochrome camera cycles through a sequence of 6 filters where the hyperspectral signature of the sensor is strongest. Once those images are taken, we can evaluate the state of the sensor, and thus the plant, based on the intensity of the hyperspectral signature we capture \cite{6080898}. 

\section{Related Work}
\subsection{Hyperspectral Imaging}
Hyperspectral imaging (HSI) has become a central technique in precision agriculture for non-destructive sensing of plant properties \cite{Bhargava2024HSIReview}. By capturing hundreds of narrow spectral bands, HSI enables fine discrimination of material and biochemical characteristics—pigments, water content, nutrient imbalances, or disease-induced stress—that traditional RGB or multi-spectral imaging cannot reveal. It supports a wide range of tasks such as water stress detection, nutrient estimation, chlorophyll monitoring, early disease detection, and yield prediction \cite{6080898, Faqir2024SoilPlantSensors}. Recent advances include lightweight HSI cameras for drones and field robots \cite{abdulridha2023stemrust}, improved calibration and illumination normalization methods, and real-time hardware acceleration using FPGAs or GPUs. However, challenges persist: high data dimensionality, illumination variability, sensor noise, and large data volumes hinder real-time field deployment \cite{Bhargava2024HSIReview}. Many recent works highlight the need for fusion with structural or proximal sensors (e.g. LiDAR or stereo cameras) to overcome these limitations.

\subsection{Agricultural Sensors}
Beyond spectral imaging, numerous non-RF proximal sensors are used in agriculture for soil and plant monitoring \cite{Thalheimer2023, Kramadhati2025-PlantSensors, Chen2018, Meshram2024SoilMoistureReview}. These include optical leaf sensors, electrochemical soil probes, and flexible wearable plant sensors. Such sensors often provide direct, localized measurements of environmental or physiological variables without the high cost or interference challenges associated with wireless RF systems. Categories and examples—Leaf-surface optical sensors: Devices like Dualex \cite{dualex} measure flavonol, anthocyanin, and chlorophyll content via leaf reflectance and fluorescence non-destructively and in real time . Soil sensors: Resistive and capacitive soil moisture probes, and emerging optical or electrochemical nitrate sensors. Flexible / wearable plant sensors: Thin, conformal sensors mounted on leaves or stems can measure humidity, transpiration, and strain \cite{Kuruppuarachchi2025WearableSensors}. Optical proximal sensing platforms: Handheld or tractor-mounted optical sensors can infer nitrogen status or canopy vigor using narrow-band reflectance or fluorescence indices. These sensors are promising but often limited in scale and durability. Integrating them with imaging or structural sensors would provide complementary data for real-time monitoring.

\subsection{Active sensors}
Active Sensors (LiDAR and Depth Sensing) Active sensors emit energy (typically laser or IR) and measure its return, allowing for 3D reconstruction of canopy structure, plant geometry, and field topography \cite{WeissBiber2011}. Among active systems, LiDAR has become the most versatile for agricultural mapping and phenotyping. Applications include canopy height modeling, biomass estimation, crop volume measurement, weed mapping, obstacle detection, and autonomous navigation for farm robots \cite{WeissBiber2011}. Recent reviews highlight both the power and challenges of LiDAR in field agriculture—especially cost, point cloud alignment, and calibration under varying light and dust conditions.

One noted method of plant sensing that is used are Radio Frequency (RF) sensors, which are attractive due to their low power nature. While advantageous in small-scale deployments, this strategy faces scalability challenges where a large amount of plant sensors must be deployed simultaneously \cite{yu2025soilmoistureRF}. This would not be feasible as the benefits of the low-power nature of RF-based sensor returns would diminish. To address this, our approach relies on using Hyperspectral Imaging (denoted HSI) as the basis for sensor interrogation. This approach removes the need for a low powered sensor on every plant, enabling us to deploy more sensors on plants in the field. 

\subsection{LiDARs in Agriculture}
Sensor Fusion and Integrated Systems—Combining multiple sensing modalities (spectral, structural, and proximal) is now recognized as a key strategy for comprehensive crop monitoring \cite{li2021imagefusion}. HSI and LiDAR fusion improves both spectral and structural accuracy, while proximal + remote sensing can scale from leaf-level precision to field-wide coverage. These multimodal systems often rely on IoT backbones and AI models for feature fusion, classification, and decision support.

\section{METHODOLOGY}
Our system is designed to automate the detection and spectral analysis of reflective plant sensors in an agricultural environment using a combination of LiDAR and adaptive optics. The overall workflow involves three main stages: (1) identifying plant sensors in the LiDAR point cloud, (2) redirecting the camera's field of view toward those sensors, and (3) capturing their hyperspectral response through a sequence of optical filters. At the core of the system is an adaptive optics assembly consisting of a LiDAR, a monochrome camera equipped with a filter wheel and liquid lens, and a Fast Steering Mirror (FSM). The LiDAR provides 3D spatial information about the environment and is used to locate potential plant sensors based on the reflectivity of the retro-reflective tape. The camera, assisted by the FSM and liquid lens, captures detailed optical data from each sensor to extract spectral information relevant to plant monitoring.

\begin{figure}
    \centering
    \includegraphics[width=1\linewidth]{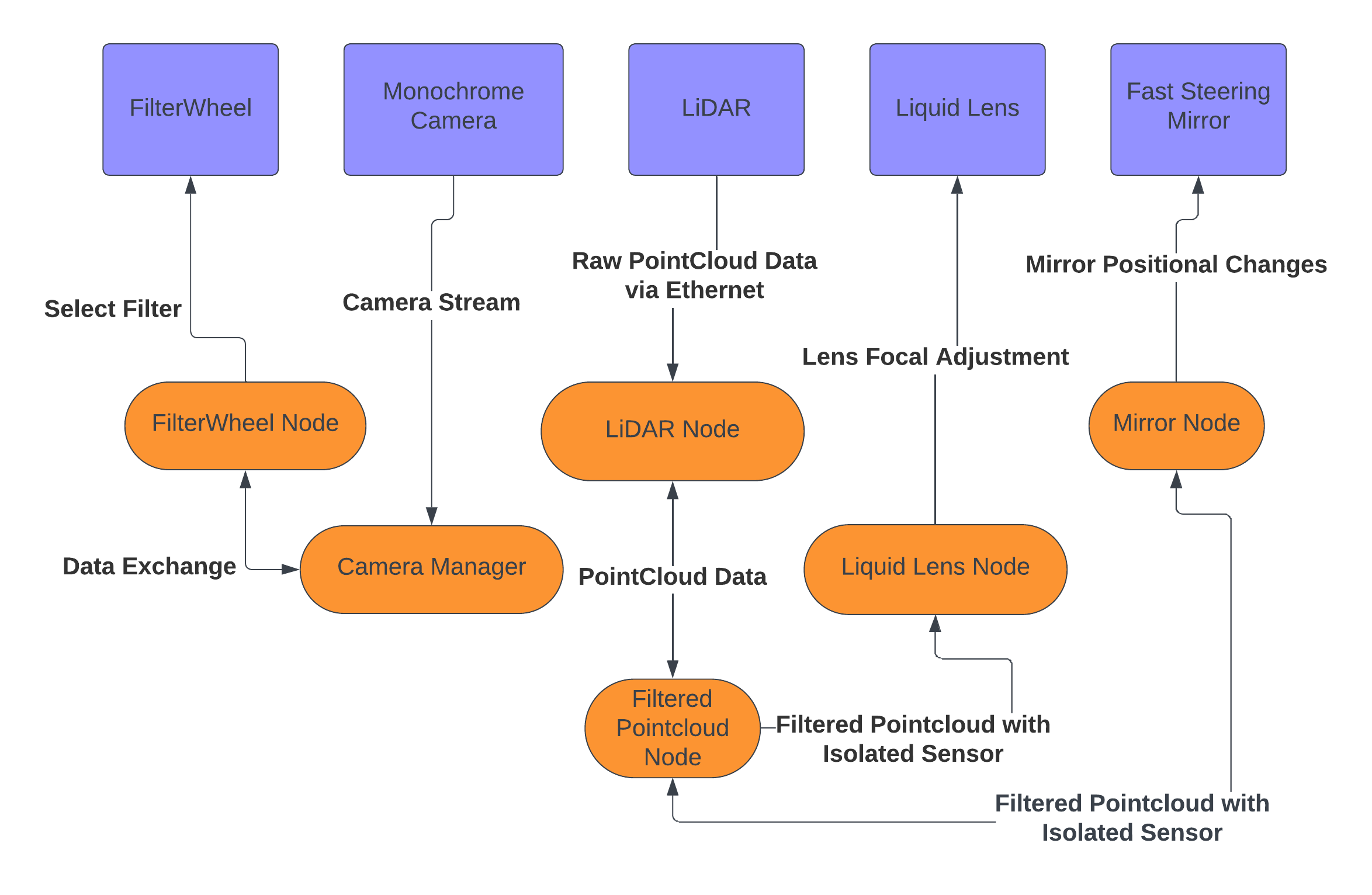}
    \caption{Legend: Purple = Hardware, Orange = ROS Node}
    \label{fig:system_outline}
\end{figure}

\subsection{Sensor Isolation in pointcloud using LiDAR Node}\label{AA}
The way this is done is we take advantage of the HDL-32 LiDAR’s ability to detect the intensity of the infrared light returned from a point in its pointcloud. Intensity is determined by the reflectivity and light absorbing properties of a surface or material. For our testing, we used retroreflective tape \cite{Ribeiro2021RetroReflective} and isolated its unique reflective properties in the pointcloud. This is done by filtering reflectivity values in the pointcloud as received from the data packets of the LiDAR in real time. The intensity to filter out was found simply via trial and error based on the intensity values in the pointcloud that our retro-reflective material appeared to be strongest. Once we had that value, we needed a way to eliminate the noise in the pointcloud that is similar in intensity to the retroreflective tape on our sensors. Surfaces like glass and other reflective materials produced spurious returns which resulted in noise in our pointcloud. We then decided to employ a filtering algorithm in order to isolate our sensors from other reflective surfaces and noise in the pointcloud. We did this using the DBSCAN k-means clustering method \cite{dbscan, clustering}. We found that this method provided a good balance in speed and accuracy when clustering the points of the retroreflective tape in the pointcloud, which gave us accurate dimensional data on where our plant sensor is in 3D space. 

\subsection{Liquid Lens Autofocus Node}
We perform the auto-focus function using the power of ROS2 nodes. The auto-focus node in our setup gets the distance of the reflective points in the LiDAR pointcloud and calculates the average of the euclidean distances of the points in that cluster. Knowing the distance of each cluster (our retroreflective tape) allows us to move the depth of field of the camera to exactly where the retroreflector would be in focus. The method we did to see where things would be in focus is by manual testing of our liquid lens up to 5 meters by seeing which diopter values are in focus within that distance. This algorithm happens perpetually while the system is on to make sure that wherever the sensor is detected in the pointcloud, the sensor is always in focus for the camera. 

\subsection{Steering Mirror Control Node}
The FSM mirror control is also done via a ROS2 node. \cite{Macenski_2022} This node listens to the Sensor Isolation node and calculates the Euclidean distance from the lidar, which is our 3D world's origin, and the Euclidean distance from the mirror itself. The FSM Node also calculates the vectors of the location of the camera, and the position of the located sensor in our pointcloud, to the location of the mirror. We calculate these vectors so that we know the position of our located sensor from the perspective of the mirror's frame. These vectors allow us to calculate a new vector that bisects the two, which is the vector toward which we want the face of our mirror to point. We construct a rotation vector that describes the movement from the default angle of the mirror to the calculated bisection and then convert that rotation to Euler angles. These angles are what we use to move the mirror so the it directs the FOV of the camera to the located plant sensor. 

\section{Hardware Components}

\subsection{Monochrome Camera - IDS Imaging U3-3270CP-M-GL} This compact, high-performance monochrome camera was chosen for its price and sensitivity to light in the NIR spectrum. It has a 60 to 50 percent quantum efficiency for light sources in the selected range we are looking for and allows adjustable exposure time so we can capture light as much light as we can from our plant sensors \cite{IDSCamera2020}.

\subsection{Filter Wheel - Thorlabs FW102C 6-Position Wheel}
This wheel holds our filters that we are using to evaluate the spectral response of our plant sensors. As mentioned earlier, we're using this 6 most responsive frequencies of the plant sensor. These frequencies are 630, 640, 650, 660, 670, and 680 nm each with a 10nm FWHM. We're controlling it via Ubuntu's Minicom interface, which allows us to send serial messages to the filter and choose which filter we would like to take a photo with \cite{ThorlabsWheel2021}. 

\begin{figure}[t]
    \centering
    \includegraphics[width=0.5\textwidth]{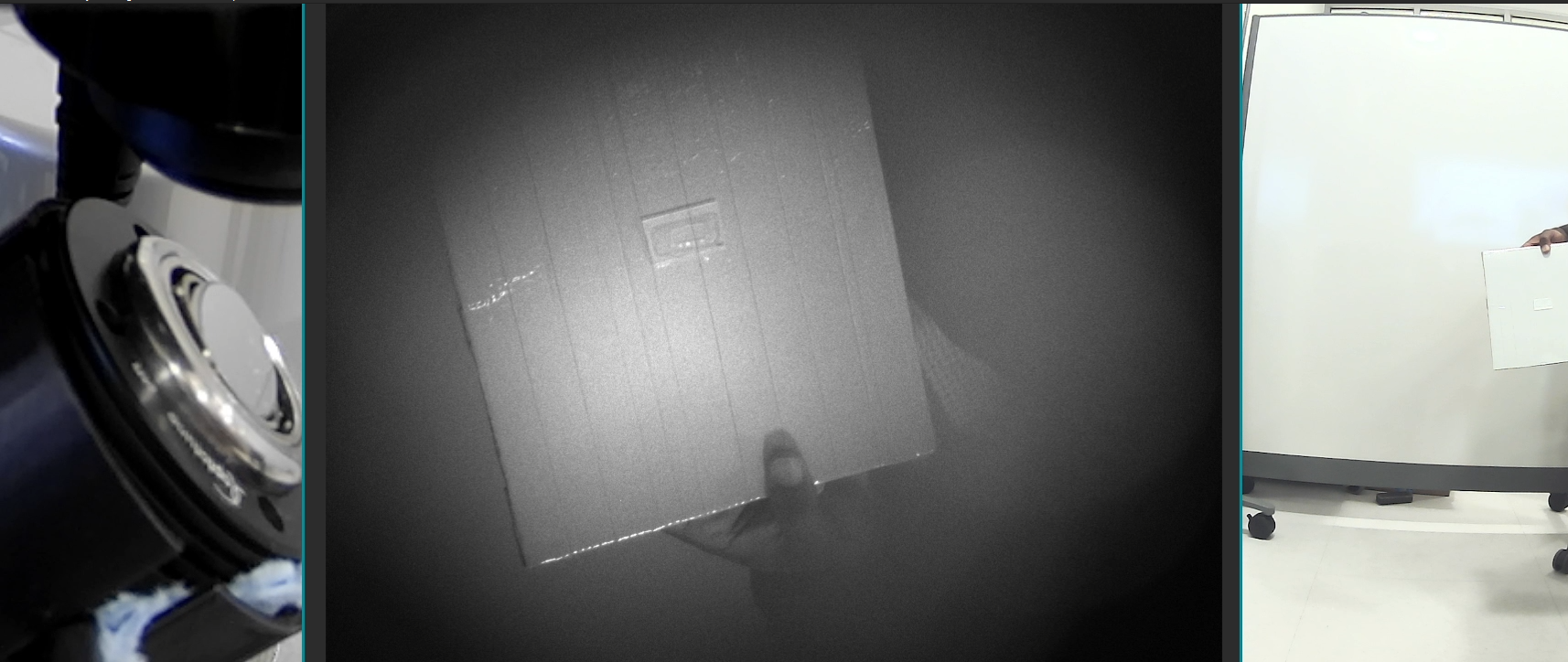}
    \caption{Sensor Tracking - Mirror tilted toward the Retro Reflective Tape with Plant Sensor attached.}
    \label{fig:sensor_tracking_right}
\end{figure}

\begin{figure}[t]
    \centering
    \includegraphics[width=0.5\textwidth]{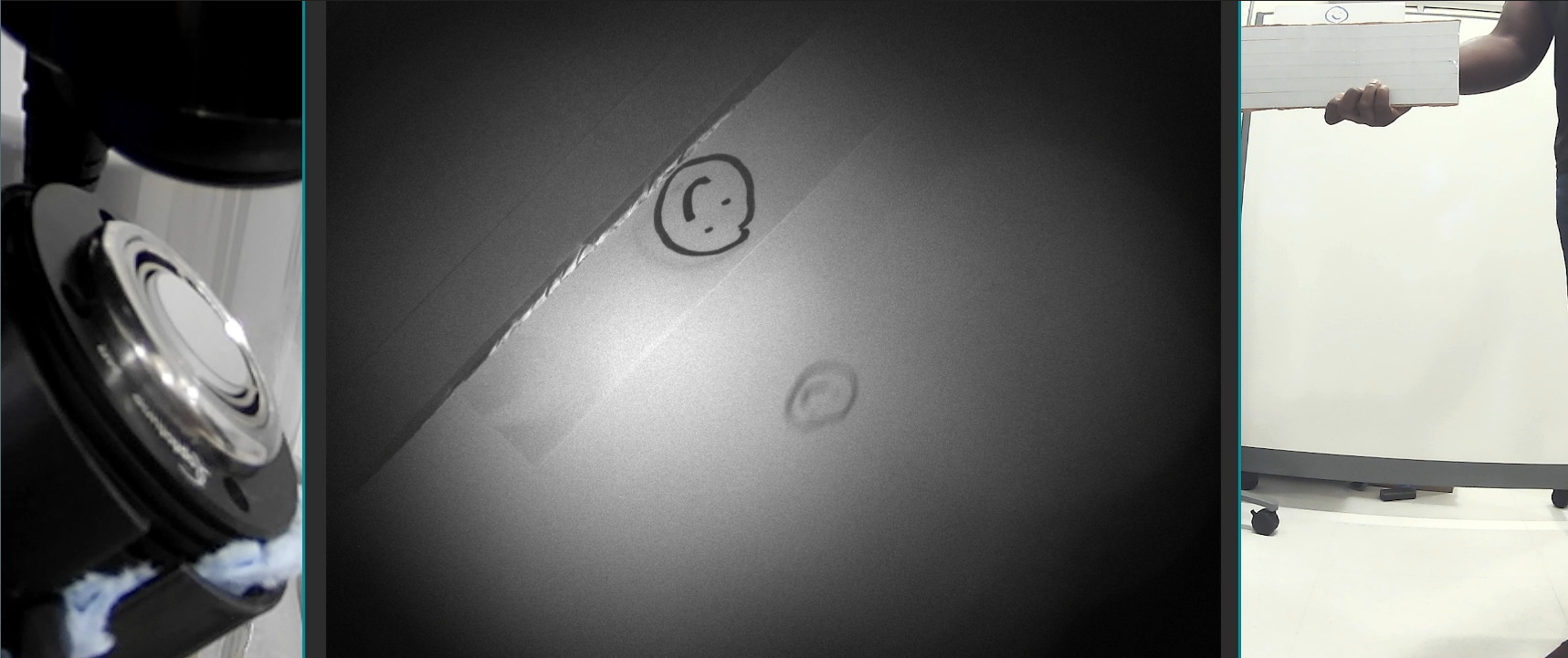}
    \caption{Demonstrating auto focus with a drawing outside of our focal range.}
    \label{fig:unfocused_smiley}
\end{figure}

\begin{figure}[t]
    \centering
    \includegraphics[width=0.5\textwidth]{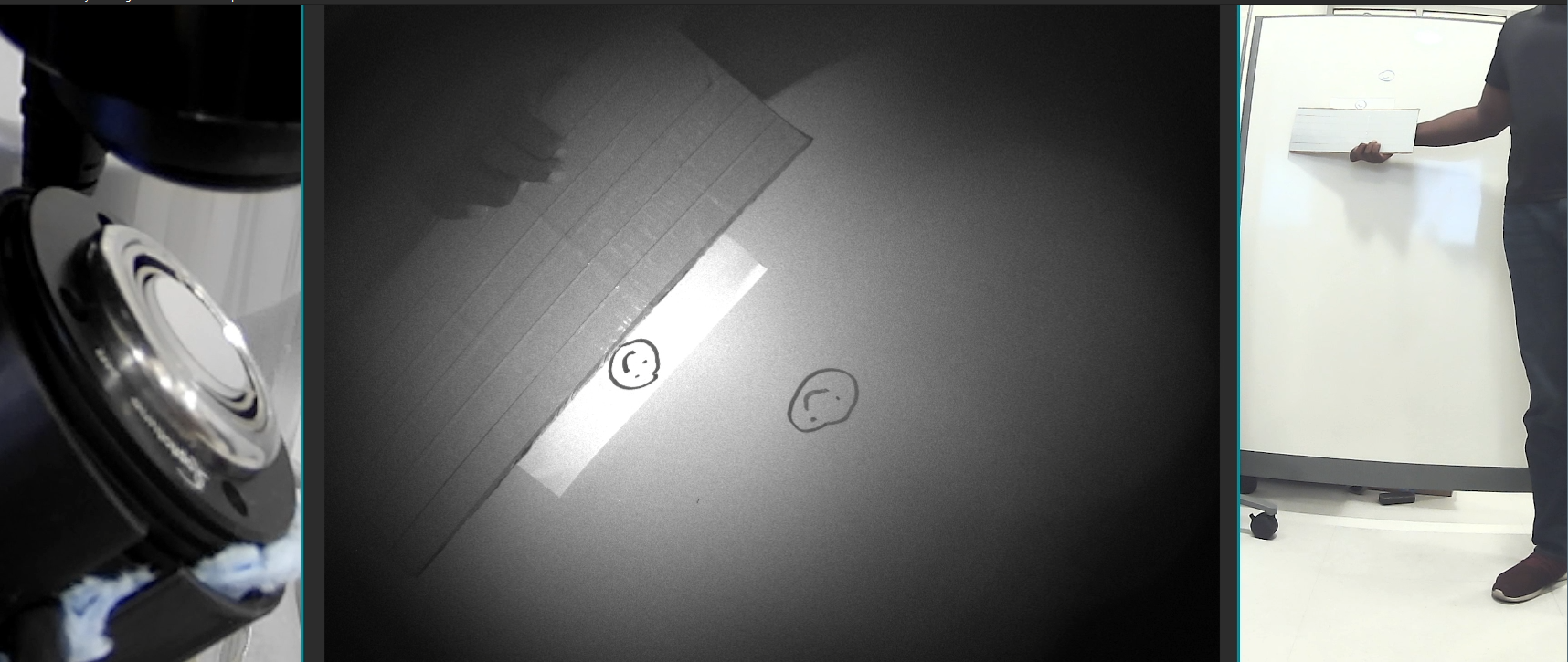}
    \caption{Demonstrating auto focus with a drawing within the focal range of our retro reflective tape.}
    \label{fig:focused_smiley}
\end{figure}

\begin{figure}[t]
    \centering
    \includegraphics[width=0.5\textwidth]{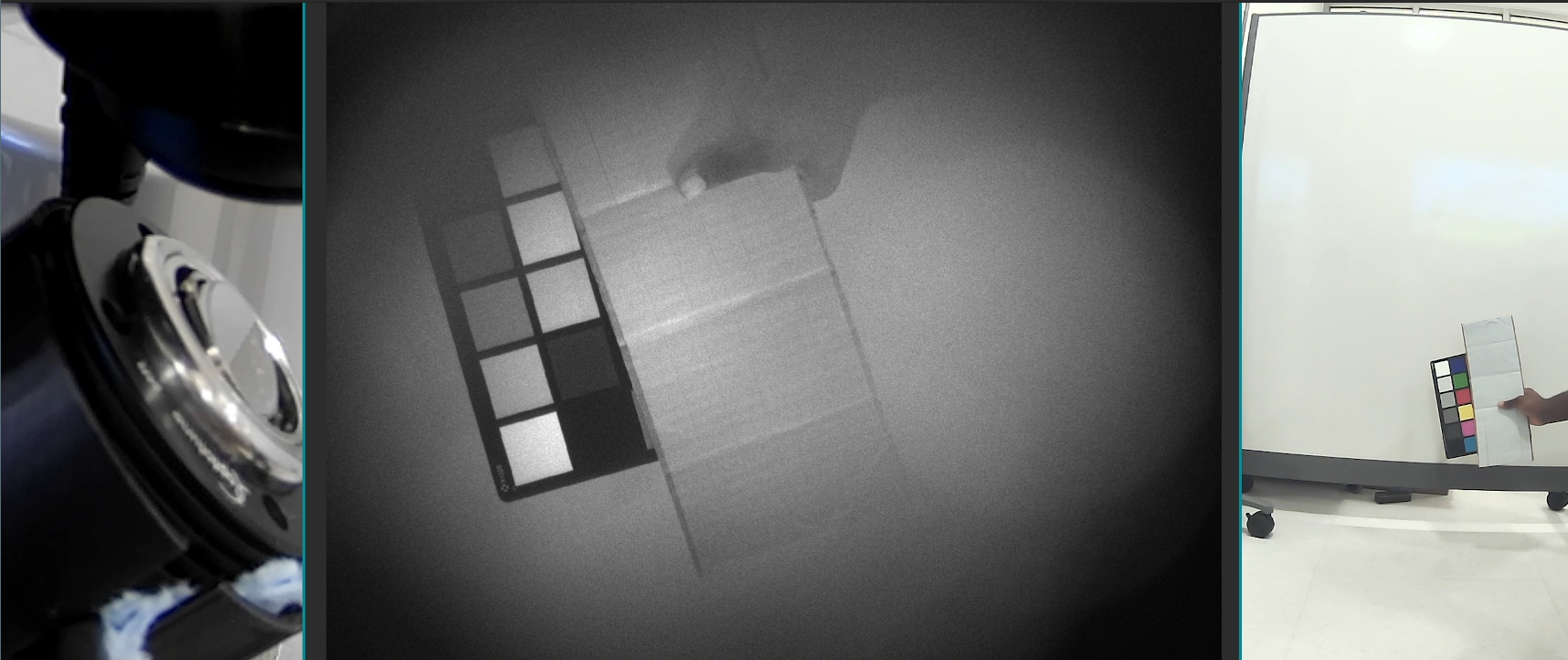}
    \caption{System Tracking Retro-Reflective Tape and color filter.}
    \label{fig:color_checking_tracking}
\end{figure}

\subsection{Liquid Lens - Optotune EL-16-40-TC}
As mentioned before, this liquid lens was chosen as a means to help focus on our targets better so that the hyperspectral response of the sensor is sharper. This liquid lens was chosen for its ability to quickly change focus based on the distance of the plant sensors, which is needed in agricultural environments where leaves and branches can move quickly because of environmental factors like the wind \cite{OptotuneLens2020}. 

\subsection{LiDAR - Velodyne HDL-32}
As mentioned before, we chose this LiDAR as our workhorse to be able to get the distance of our sensors quickly while also being able to isolate them from other objects and noise in the field of view of our camera system. It has 32 layers of vertical beams, which we found is just enough to be able to get reliable segmentation between our plant sensors and environment noise \cite{Velodyne2020}. 

\subsection{Fast Steering Mirror(FSM) - Optotune MR-15-30-PS Silver Coated Mirror}
This Fast Steering mirror was chosen as we desired for our camera's FOV to be able to adjust both pitch and yaw, with the least amount of moving parts as possible. We also required it to have a silver coating so that as much of the light from the acquired plant sensor is received into the camera as possible with the least amount of color distortion as possible \cite{OptotuneMirror2020}. 

\section{Results}

\subsection{Point Cloud Segmentation}
As can be seen in the images and attached videos, we were able to successfully isolate our sensor in 3D space. We are able to do this from between 0.8 to 2 meters in front of our LiDAR. Anything further, the LiDAR cannot pick up. We believe this is due to the nature of the Avalanche Photodiodes (APD) within the HDL-32 LiDAR, which is not as light sensitive as modern SPAD based LiDARS. Regardless, we were able to achieve full isolation. 

\subsection{FSM - Plant Sensor Acquisition}
As shown in figure \ref{fig:sensor_tracking_right}, we can see images of the mirror locking onto our plant sensor when it is within the field of view that the mirror is capable of viewing. The limitation of the FSM mirror is 39 degrees in any direction from its default normal orientation. So for this experiment we are assuming our plant sensors to be within 39 degrees of the front of the FSM.

\subsection{Color Segmentation}
In the referenced image \ref{fig:color_checking_tracking}, we can see that the filters are isolating the colors in the 630 to 680 nm range as described. We can see that the reds and red-adjacent colors are visibly brighter in our photos than other colors.

\subsection{Focus and De-focus}
In figures \ref{fig:unfocused_smiley} and \ref{fig:focused_smiley}, we show that our system successfully focuses and de-focuses based on how far our detected sensor is from the LiDAR. As mentioned before, this is a very important feature so that our system can accurately and effectively receive as much possible light from the detected sensor.

\section{Conclusions}
We present a proof-of-concept system integrating passive, metasurface-based leaf sensors with a custom LiDAR and optics system for accurately detecting plant sensors. While the results shown are promising, we have recognized some limitations that should be mentioned and recognized and some potential solutions and future plans for the system. We did mention earlier that we're using a LiDAR with APD's and not a modern SPAD-based lidars. Another problem is that the specific LiDAR that we're using cant see our sensor's retro reflective tape past 2 meters currently. Where the two ideas connect is that we may be able to view and isolate our sensors from further distances with more modern LiDARs but the system may end up being more expensive. Additionally, noise could still potentially be a big problem depending on where the system is implemented. We have designed our system in mind of agricultural contexts outdoors but with the rise of indoor and vertical farming methods where crops may be surrounded by reflective and metallic materials indoors, our solution may struggle to keep plant sensors isolated from its environment. But in situations like this in the future, we can work on refining our isolation algorithm, and maybe that, combined with modern LiDARs, will be able to help isolate sensors not just in outdoor agricultural environments. 

\bibliographystyle{IEEEtran}


\end{document}